\documentclass[onecolumn,doublespaced, 12pt,journal]{IEEEtran}

\usepackage{amsfonts}
\usepackage{trsym}
\usepackage{graphicx}
\usepackage{psfrag}
\usepackage{epsfig}
\usepackage{bm}
\newcommand{\mb}{\mathbf}
\newcommand{\mr}{\mathrm}

 \makeatletter
\let\IEEEproof\proof
\let\IEEEendproof\endproof
\let\proof\@undefined
\let\endproof\@undefined
\makeatother

\usepackage{amsmath,amsthm,amscd,amssymb} 

\let\proof\IEEEproof
\let\endproof\IEEEendproof

\newtheorem*{lemma1}{Lemma 1}
\newtheorem*{lemma2}{Lemma 2}
\newtheorem*{lemma3}{Lemma 3}

\newtheorem*{theorem1}{Theorem 1}
\newtheorem*{theorem2}{Theorem 2}
\newtheorem*{corollary1}{Corollary 1 of Theorem 1}

\title{On the Distortion of the Eigenvalue Spectrum in MIMO Amplify-and-Forward Multi-Hop Channels}

\author{J\"{o}rg Wagner and Armin Wittneben\\
Communication Technology Laboratory, ETH Zurich
\\Sternwartstrasse 7, 8092 Zurich, Switzerland \\
\{wagner, wittneben\}@nari.ee.ethz.ch}

\begin{document}




\maketitle
\begin{abstract}
 Consider a wireless MIMO
multi-hop channel with \mbox{$n_\mr{s}$} non-cooperating source
antennas and \mbox{$n_\mr{d}$} fully cooperating destination
antennas, as well as \mbox{$L$} clusters containing \mbox{$k$}
non-cooperating relay antennas each. The source signal traverses
all \mbox{$L$} clusters of relay antennas, before it reaches the
destination. When relay antennas within the same cluster scale
their received signals by the same constant before the
retransmission, the equivalent channel matrix \mbox{$\mb{H}$}
relating the input signals at the source antennas to the output
signals at the destination antennas is proportional to the product
of channel matrices \mbox{$\mb{H}_l$}, \mbox{$l=1,\ldots,L+1$},
corresponding to the individual hops. We perform an asymptotic
capacity analysis for this channel as follows: In a first instance
we take the limits \mbox{$n_\mr{s}\rightarrow\infty$},
\mbox{$n_\mr{d}\rightarrow\infty$} and
\mbox{$k\rightarrow\infty$}, but keep both
\mbox{$n_\mr{s}/n_\mr{d}$} and \mbox{$k/n_\mr{d}$} fixed. Then, we
take the limits \mbox{$L\rightarrow\infty$} and
\mbox{$k/n_\mr{d}\rightarrow\infty$}. Requiring that the
$\mb{H}_l$'s satisfy the conditions needed for the
Mar\v{c}enko-Pastur law, we prove that the capacity scales
linearly in \mbox{$\min\{n_\mr{s},n_\mr{d}\}$}, as long as the
ratio $k/n_\mr{d}$ scales at least linearly in $L$. Moreover, we
show that up to a noise penalty and a pre-log factor the capacity
of a point-to-point MIMO channel is approached, when this scaling
is slightly faster than linear. Conversely, almost all spatial
degrees of freedom vanish for less than linear scaling.
\end{abstract}

\newpage

\section{Introduction}
We consider coherent wireless multiple-input-multiple-output
(MIMO) communication between $n_\mr{s}$ non-cooperating source
antennas and \mbox{$n_\mr{d}$} fully cooperating destination
antennas. In this paper it is assumed that the source antennas are
either far apart or shadowed from the destination antennas. The
installation of intermediate nodes that relay the source signals
to the destination (multi-hop) is well known for being an
efficient means for improving the energy-efficiency of the
communication system in this case. In the resulting network, the
signals traverse \mbox{$L$} clusters containing \mbox{$k$} relay
antennas each, before they reach the destination. Generally,
signals transmitted by the source antennas might not only be
received by the immediately succeeding cluster of relay antennas,
but possibly also by clusters that are farther away or by the
destination. While such receptions could well be exploited for
achieving higher transmission rates, we assume them to be strongly
attenuated and ignore them in this paper.

In the most basic MIMO multi-hop network architecture, the relay
antennas in the clusters do not cooperate. Since non-cooperative
decoding of the interfering source signals at the individual relay
antennas drastically reduces the achievable rate in the network, a
simple amplify-and-forward operation becomes the relaying strategy
of choice. That is, at each antenna a scaling of the received
signals by a constant is performed before the retransmission.
While this approach is cheap in terms of computational complexity,
and also does not require any channel-state information at the
relay nodes, it clearly suffers from noise accumulation. This
basic network has been studied extensively by Borade, Zheng and
Gallager for independent identically distributed (i.i.d.) Rayleigh
fading channel matrices. In references
\cite{boradeZhengGallager03,boradejournal} they showed for
\mbox{$n\triangleq n_\mr{s}=n_\mr{d}=k$} that all \mbox{$n$}
spatial degrees of freedom are available in this network for a
fixed $L$ at high signal-to-noise ratio (SNR). More generally,
they also showed that all degrees of freedom are available, if $L$
as a function of the SNR fulfills
\mbox{$\lim_{\mathsf{snr}\rightarrow\infty}L(\mathsf{snr})/\log
\mathsf{snr}=0$}.

While this result gives a design criterion how the SNR should be
increased with the number of hops in the network, it does not give
any insights into the eigenvalue distribution of the product of
random matrices $\mb{C}$ specifying the mutual information between
input and output of the vector channel. For fixed $L$, only recently
this eigenvalue distribution has been characterized in the large
antenna limit \cite{leveque}. Based on a theorem from large random
matrix theory \cite{silverstein}, the authors showed that it
converges to a deterministic function, and gave a recursive formula
for the corresponding Stieltjes transform. Moreover, the reference
reports that the asymptotic eigenvalue distribution of $\mb{C}$,
which is in fact the product of the signal covariance matrix
$\mb{R}_s$ and inverse noise covariance matrix $\mb{R}_n^{-1}$ at
the destination, approaches the Mar\v{c}enko-Pastur law in the large
dimensions limit for \mbox{$\beta_r\triangleq
k/n_\mr{d}\rightarrow\infty$}, but \mbox{$\beta_s\triangleq
n_\mr{s}/n_\mr{d}$} and \mbox{$L$} fixed. Since the
Mar\v{c}enko-Pastur law is also the limiting eigenvalue distribution
of the classical point-to-point MIMO channel, this means that up to
a noise penalty and a pre-log factor the point-to-point capacity is
approached in this case.

By considering the limiting eigenvalue distributions of the signal
and noise covariance matrices separately, we are able to
generalize this result for the case $L\rightarrow \infty$ in this
paper. In essence, we show that $\beta_r$ needs to grow at least
linearly with $L$ in order to sustain a non-zero fraction of the
spatial degrees of freedom in the system, i.e., linear capacity
scaling in \mbox{$\min\{n_\mr{s},n_\mr{d}\}$}. Moreover, when the
scaling is faster than linear, the limiting eigenvalue
distribution of $\mb{C}$ is given by the Mar\v{c}enko-Pastur law.
That is, we are able exploit the spatial degrees of freedom
without increasing the SNR at the receiver at the expense of
employing more relay antennas. Returning to the result by Borade
et al., where degrees of freedom are sustained by increasing the
SNR, according to our result the number of relays per layer can be
seen as a second resource besides the transmit power for
compensating the capacity loss in the multi-hop network.

Another contribution of this paper lies in bridging the gap
between the results obtained by M\"{u}ller in reference
\cite{mueller2} on the one hand and by Morgenshtern and
B\"{o}lcskei in references \cite{venjamin,veniajournal} on the
other hand. In the first reference, it is shown in the large
dimensions limit that almost all singular values of a product of
independent random matrices fulfilling the conditions needed for
the Mar\v{c}enko-Pastur law go to zero as the number of
multipliers grows large, while the aspect ratios of the matrices
are kept finite. This implies that almost all spatial degrees of
freedom in a MIMO amplify-and-forward multi-hop network as
described above vanish as \mbox{$L$} goes to infinity. On the
other hand, \cite{venjamin,veniajournal} were the first papers
which proved in the large dimensions limit (for $L=1$) that the
capacity of a point-to-point MIMO link is approached up to a noise
penalty and a pre-log factor, if \mbox{$\beta_r\rightarrow
\infty$} and \mbox{$\beta_s$} is kept fixed. In \cite{boelcskei}
the same result had been proven for the less general case that
$n_\mr{s}$ and $n_\mr{d}$ are fixed and $k\rightarrow\infty$. The
mechanisms discovered in these papers apparently act as antipodal
forces with respect to the limiting eigenvalue distributions of
products of random matrices. While increasing the number of hops
distorts this distribution in an undesired fashion, increasing the
ratio between the number of relays and destination antennas allows
for recovering the original distribution corresponding to a
point-to-point channel. In this paper, we answer the question how
these two effects can be balanced, i.e., how fast must
\mbox{$\beta_r$} grow with \mbox{$L$} in order to sustain a
non-zero fraction of spatial
degrees of freedom as \mbox{$L$} grows without bounds. 

\section{Notation}
The superscripts $^H$ and $^*$ stand for conjugate transpose and
complex conjugate, respectively. $\textsf{E}_A$ denotes the
expectation operator with respect to the random variable $A$.
$\det(\mb{A})$, $\mr{Tr}(\mb{A})$ and $\lambda_i\{\mb{A}\}$ stand
for determinant, trace and the $i$th eigenvalue of the matrix
$\mb{A}$. $a(i)$ is the $i$th element of the vector $\mb{a}$.
Throughout the paper all logarithms, unless specified otherwise,
are to the base $e$. $\|\mb{a}\|$ denotes the Euclidean norm of
the vector $\mb{a}$, $\|\mb{A}\|_\mr{Tr}$ the Trace norm of the
matrix $\mb{A}$. By $\textsf{Pr}[A]$ we denote the probability of
the event $A$.

Furthermore, we use the standard \mbox{${\cal O}(\cdot),
\Omega(\cdot), \Theta (\cdot)$} notations for characterizing the
asymptotic behavior of some function \mbox{$f(\cdot)$} according
to
\begin{align}
&f(n)\in {\cal O}(g(n)) \textrm{ if } \exists  M, n_0>0:
M|g(n)|>|f(n)|, \forall n\geq n_0,\nonumber\\
&f(n)\in \Omega(g(n)) \textrm{ if } \exists  M , n_0>0:
M|g(n)|<|f(n)|, \forall n>n_0,\nonumber\\
&f(n)\in \Theta (g(n)) \textrm{ if } f(n)\in {\cal O}(g(n))
\textrm{ and } f(n)\in \Omega (g(n)).\nonumber
\end{align}

Finally, we define the function $1\{x\}$ to be 1 if $x$ is true
and zero otherwise. $\delta(x)$ and $\sigma(x)$ denote Dirac delta
and Heaviside step function, respectively.

\section{Transmission Protocol \& System Model}
\begin{figure}
        \centering
        \psfrag{H1}{$\mb{H}_1$}
        \psfrag{HL}{$\mb{H}_{L}$}
        \psfrag{HKp1}{$\mb{H}_{L+1}$}

        \includegraphics[width=\columnwidth]{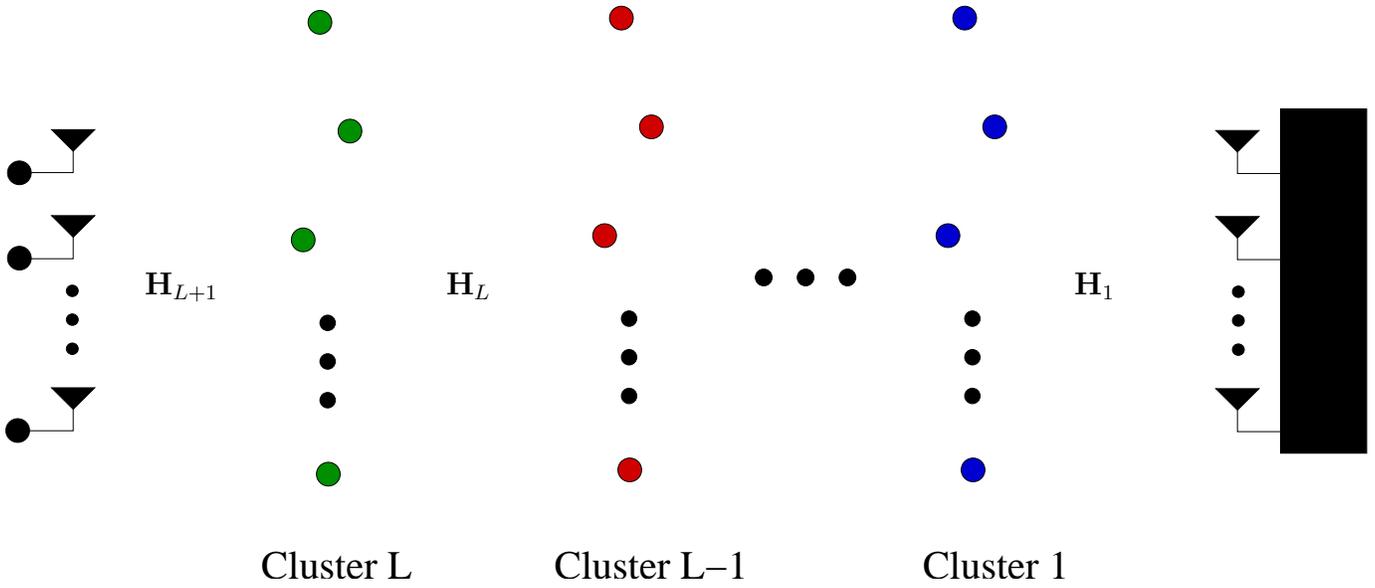}
        \caption{\mbox{$n_\mr{s}$} non-cooperating source antennas transmit to a destination terminal with \mbox{$n_\mr{d}$} antennas via \mbox{$L$} clusters of $k$ non-cooperating relay antennas.}
        \label{fig: setup}
\end{figure}

We label the clusters of relay antennas by \mbox{$\mathcal{C}_1,
\ldots, \mathcal{C}_L$}. Cluster $\mathcal{C}_1$ denotes the one
next to the destination, $\mathcal{C}_L$ the one next to the
sources (refer to Fig. \ref{fig: setup}). We assume the
\mbox{$L+1$} single hop channels between sources, relay clusters
and destinations to be frequency-flat fading over the bandwidth of
interest, and divide the transmission into $L+1$ time slots. In
time slot $T=1$ the sources transmit to \mbox{$\mathcal{C}_L$}.
The transmission is described by the transmit vector
\mbox{$\mb{s}\in\mathbb{C}^{n_\mr{s}}$}, the matrix
\mbox{$\mb{H}_{L+1}\in\mathbb{C}^{k\times n_\mr{s}}$},
representing the vector channel between sources and
\mbox{$\mathcal{C}_L$}, the vector $\mb{n}_{L}\in\mathbb{C}^{k}$,
representing the receiver front-end noise introduced in
\mbox{$\mathcal{C}_L$}, the receive vector
\mbox{$\mb{y}_L\in\mathbb{C}^{k}$} and the linear mapping
\begin{align}
\mb{y}_L=\mb{H}_{L+1}\mb{s}+\mb{n}_{L}.\nonumber
\end{align}
Here and also in the subsequent equations, the $i$th elements of
the transmit, receive and noise vectors correspond to the $i$th
antenna in the respective network stage.

The time slots \mbox{$T=\{2,\ldots,L\}$} are used for relaying the
signals from cluster to cluster. In time slot $T$ the relay
antennas in $\mathcal{C}_{L-T+2}$ transmit scaled versions of the
signals received in time slot $T-1$ to $\mathcal{C}_{L-T+1}$. That
is, with $l=L-T+2$ the transmit vector of the $l$th relay cluster
$\mb{r}_l\in\mathbb{C}^k$ is computed from the respective receive
vector $\mb{y}_l\in\mathbb{C}^k$ according to
\begin{align}
\mb{r}_l=\sqrt\frac{\alpha_l}{k}\mb{y}_l\nonumber,
\end{align}
where $\mb{\alpha}_l\in\mathbb{R}$ is a cluster specific constant
of proportionality specifying the ratio between receive and
transmit power. The transmission in time slot $T$ is then
described by
\begin{align}
\mb{y}_{l-1}=\mb{H}_{l}\mb{r}_l+\mb{n}_{l-1}.\nonumber
\end{align}
Here, \mbox{$\mb{H}_{l}\in\mathbb{C}^{k\times k}$} represents the
channel between \mbox{$\mathcal{C}_l$} and
\mbox{$\mathcal{C}_{l-1}$}, $\mb{n}_{l-1}\in\mathbb{C}^{k}$ the
receiver front-end noise introduced in \mbox{$\mathcal{C}_{l-1}$},
and \mbox{$\mb{y}_{l-1}\in\mathbb{C}^{k}$} is the corresponding
receive vector. Thus, the signals traverse one hop per time slot.
In time slot \mbox{$T=L+1$}, \mbox{$\mathcal{C}_{1}$} finally
forwards its received signals to the destination. Again, the
transmit vector is computed according to
$\mb{r}_1=\sqrt{\frac{\alpha_1}{k}}\mb{y}_1$. Denoting the matrix
representing the channel between \mbox{$\mathcal{C}_{1}$} and the
destination by $\mb{H}_1$, the vector representing the receiver
front-end noise at the destination by
\mbox{$\mb{n}_{d}\in\mathbb{C}^{n_\mr{d}}$} and the receive vector
by $\mb{y}\in\mathbb{C}^{n_\mr{d}}$ this transmission is described
by
\begin{align}
\mb{y}=\mb{H}_{1}\mb{r}_1+\mb{n}_\mr{d}.\nonumber
\end{align}
Putting everything together, the input-output relation of the
channel as seen from source to destination antennas over $L+1$
time slots can be written as
\begin{align}\nonumber
&\mb{y}=
\sqrt{\frac{\prod_{l=1}^L\alpha_l}{k^L}}\mb{H}_{1}\cdots\mb{H}_{L+1}
\mb{s}+
\mb{n}_{\mr{d}}+\sum_{l=1}^{L}\sqrt{\frac{\prod_{l^{'}=1}^l\alpha_{l^{'}}}{k^{l}}}\mb{H}_{1}\cdots\mb{H}_{l}
\mb{n}_{l}.
\end{align}
We model the entries of all noise vectors as zero-mean circular
symmetric complex Gaussian random variables of unit-variance that
are white both in space and time. The channel matrices are
independent and their elements are assumed to be i.i.d. zero-mean
random variables of unit-variance. Moreover, we impose the per
antenna power constraints
\mbox{$\textsf{E}[s(i)s(i)^*]=P/n_\mr{s}$} and
\mbox{$\textsf{E}[r_l(i)r_l(i)^*]= P/k$} for
\mbox{$l=\{1,\ldots,L\}$}. The relay antenna power constraints are
fulfilled, if the scaling factors $\alpha_i$ satisfy
$\alpha\triangleq\alpha_1=\ldots=\alpha_L=P/(P+1)$.

\section{Ergodic Capacity \& Convergence of Eigenvalues}
While full cooperation and the presence of full channel-state
information is assumed at the destination antennas, source and
relay antennas do not cooperate and also do not possess any
channel-state information. Under these assumptions, the ergodic
mutual information $I(\mb{s};\mb{y})$ is maximized, when the
entries of $\mb{s}$ are zero-mean circularly symmetric complex
Gaussian random variables of variance $P/n_\mr{s}$ that are white
over both space and time \cite{telatar95capacity}. For this input
distribution, $I(\mb{s};\mb{y})$ is fully characterized by the
joint probability distribution of the eigenvalues of the product
\mbox{$\mb{C}=\mb{R}_s\mb{R}_n^{-1}$}, where
$\mb{R}_s\in\mathbb{C}^{n_\mr{d}\times n_\mr{d}}$ and
$\mb{R}_n\in\mathbb{C}^{n_\mr{d}\times n_\mr{d}}$ denote the
signal and noise covariance matrices at the destination of the
multi-hop channel. These covariance matrices are given by
\begin{align}
\mb{R}_s&=\textsf{E}_\mb{s}\left[\frac{\alpha^L}{k^L}\mb{H}_{1}\cdots\mb{H}_{L+1}
\mb{s}\mb{s}^\mr{H}\mb{H}_{L+1}^\mr{H}\cdots\mb{H}_{1}^\mr{H}\right]\nonumber\\
&=\frac{P\alpha^L}{n_\mr{s}k^L}\mb{H}_{1}\cdots\mb{H}_{L+1}\mb{H}_{L+1}^\mr{H}\cdots\mb{H}_{1}^\mr{H},\nonumber\\
\mb{R}_n&=\textsf{E}_{\mb{n}_\mr{d},\mb{n}_1,\ldots,\mb{n}_L}\left[\left(\mb{n}_{\mr{d}}+\sum_{l=1}^{L}\sqrt{\frac{\alpha^{l}}{k^{l}}}\mb{H}_{1}\cdots\mb{H}_{l}
\mb{n}_{l}\right)\left(\mb{n}_{\mr{d}}+\sum_{l=1}^{L}\sqrt{\frac{\alpha^l}{k^{l}}}\mb{H}_{1}\cdots\mb{H}_{l}
\mb{n}_{l}\right)^\mr{H}\right]\nonumber\\
&=\mb{I}_{n_\mr{d}}+\sum_{l=1}^{L}\left(\frac{\alpha}{k}\right)^l\mb{H}_{1}\cdots\mb{H}_{l}\mb{H}_{l}^\mr{H}\cdots\mb{H}_{1}^\mr{H}.\nonumber
\end{align}
We define the the empirical eigenvalue distribution (EED) of the
matrix $\mb{A}$ as
\begin{align}
F_{\mb{A}}^{(n)}(x)=\frac{1}{n}\sum_{i=1}^{n}
1\{\lambda_i\{\mb{A}\}<x\}.
\end{align}
With this notation the ergodic capacity of the multi-hop channel
in nats per channel use is obtained as
\begin{align}
C&=\frac{1}{L+1}\cdot\textsf{E}_{\mb{C}}\left[\log\det\left(\mb{I}_\mr{d}+\mb{C}\right)\right]\nonumber\\
&=\frac{1}{L+1}\cdot\textsf{E}_{\mb{C}}\left[\sum_{i=1}^{n_\mr{d}}\log(1+\lambda_i\{\mb{C}\})\right]\nonumber\\
&=\frac{1}{L+1}\cdot\textsf{E}_{\mb{C}}\left[\int_0^\infty
\log(1+x)\cdot dF_{\mb{C}}^{(n_\mr{d})}(x)\right]\label{eq:
capacity}.
\end{align}
Note that the pre-log factor \mbox{$(L+1)^{-1}$} accounting for
the use of \mbox{$L+1$} time slots can be lowered by initiating
the next source antenna transmission after $L_0<L$ time slots.
From a practical perspective $L_0$ needs to be chosen large
enough, such that the interference imposed on the previously
transmitted signal is negligible. It is important to have this
fact in mind whenever we take the limit $L\rightarrow \infty$,
which formally drives the ergodic capacity to zero. In this paper
we are interested in the scaling of the capacity in the number of
source and destination antennas. Accordingly, we focus on the case
where both these quantities grow large. From \cite{leveque} we
know that $F_{\mb{C}}^{(n)}(x)$ converges almost surely (a.s.) to
some asymptotic distribution \mbox{$F_{\mb{C}}(x)$}, when
\mbox{$n_\mr{s}\rightarrow\infty$},
\mbox{$n_\mr{d}\rightarrow\infty$} and
\mbox{$k\rightarrow\infty$}, but the ratios \mbox{$\beta_s=
n_\mr{s}/n_\mr{d}$} and \mbox{$\beta_r = k/n_\mr{d}$} are fixed.
Here, we mean by the convergence of an EED $F_{\mb{A}}^{(n)}(x)$
to some deterministic function $F_{\mb{A}}(x)$ that
\begin{align}\nonumber
&\mathsf{Pr}\left[\lim_{n\rightarrow \infty} \sup_{x\in\mathbb{R}}
|F_{\mb{A}}^{(n)}(x)-F_{\mb{A}}(x)|=0\right]=1.
\end{align}
We will refer to the density \mbox{$f_{\mb{A}}(x)=\frac{d}{dx}
F_{\mb{A}}(x)$} as the limiting spectral measure (LSM)
subsequently.

Returning to the capacity expression \eqref{eq: capacity}, we can
infer that for \mbox{$n_\mr{s}\rightarrow\infty$},
\mbox{$n_\mr{d}\rightarrow\infty$} and
\mbox{$k\rightarrow\infty$}, and \mbox{$\beta_s$} and
\mbox{$\beta_r $} fixed, $C$ converges to the quantity $C_\infty$
defined as
\begin{align}
C_{\infty}&\triangleq\frac{1}{L+1}\cdot\int_0^{\infty}\log(1+x)\cdot
f_{\mb{C}}(x)\cdot dx\nonumber.
\end{align}

\section{Preliminaries From Random Matrix Theory}

We briefly repeat some preliminaries from the theory of large random matrices subsequently.

\subsection{Stieltjes Transform}
We define the Stieltjes transform of some LSM \mbox{$f(\cdot)$} as
\begin{align}
G(s)\triangleq\int_{-\infty}^\infty \frac{f(x)}{s+x}\cdot dx.
\end{align}
We stick to the definition in \cite{mueller2} here, while it is
generally more common to define the Stieltjes transform with a
minus sign in the denominator. The LSM is
uniquely determined by its Stieltjes transform. 
 While it is impossible to
find a closed form expression for the LSM of a random matrix in many
cases, implicit (polynomial) equations for the corresponding
Stieltjes transform can sometimes be obtained. Accordingly, the
Stieltjes transform plays a prominent role in large random matrix
theory. A transform pair appearing again and again in the course of
this paper is the following:
\begin{align}
f(x)=\delta(x-x_0) \TransformHoriz G(s)=\frac{1}{s-x_0}.
\end{align}

\subsection{Mar\v{c}enko-Pastur Law}
The result presented in this paper is valid for the class of
random matrices fulfilling the conditions for the
Mar\v{c}enko-Pastur law \cite{Marcenko,bai}, which we briefly
repeat here. Let \mbox{$\mb{X}\in\mathbb{C}^{k_0\times k_1}$} be a
random matrix whose entries are i.i.d. zero-mean distributed and
of unit-variance. If both \mbox{$k_0\rightarrow \infty$} and
\mbox{$k_1\rightarrow \infty$}, but \mbox{$\beta=k_1/k_0$} is kept
finite, then the Stieltjes transform of the LSM of
\mbox{$\frac{1}{k_1}\mb{X}\mb{X}^\mr{H}$} is given by
\begin{align}\label{eq: Marcenko}
G_\mr{MP}^{(\beta)}(s)&=\frac{\beta^{-1}-1-s\pm
\sqrt{s^2+2\left(\beta^{-1}+1\right)s+\left(\beta^{-1}-1\right)^2}}{2s\beta^{-1}}.
\end{align}
The corresponding LSM can be written in closed form.

\subsection{Concatenated Vector Channel}
Our result is based on a theorem on the concatenated vector
channel proven in \cite{mueller2} using the S-transform
\cite{stransform,muller-random}: \noindent
\begin{theorem1}Let \mbox{$\mb{M}_1\in\mathbb{C}^{k_0\times
k_1},\ldots,\mb{M}_{N}\in\mathbb{C}^{k_{N-1}\times k_{N}}$} be
independent random matrices fulfilling the conditions for the
Mar\v{c}enko-Pastur law, whose elements are of unit-variance, and
define \mbox{$\beta_n=\frac{k_n}{k_0}$}. Then the Stieltjes
transform of the LSM corresponding to \mbox{$1/\left(k_1\cdots
k_{N}\right)\mb{M}_1\cdots\mb{M}_N\mb{M}_N^\mr{H}\cdots\mb{M}_1^\mr{H}$}
fulfills the implicit equation
\begin{align}\label{eq: concatenated stieltjes}
\frac{G(s)}{\beta_{N}}\prod_{n=0}^{N-1}\frac{sG(s)-1+\beta_{n+1}}{\beta_n}+sG(s)&=1.
\end{align}
\end{theorem1}
\noindent Note that we normalize with respect to \mbox{$k_N$}
rather than with respect to \mbox{$k_0$} as done in
\cite{mueller2}. The Stieltjes transform \mbox{$\tilde{G}(s)$}
therein relates to $G(s)$ as \mbox{$G(s)=\beta_N \tilde{G}(\beta_N
s)$}.

\section{Capacity Scaling}
We formalize our result in the subsequent theorem. It is important
to note that taking the limits in the LSM of a random matrix means
that the dimensions of this matrix are already taken to infinity.
For the case at hand that is, we first take the limits
$n_\mr{s}\rightarrow \infty$, $n_\mr{d}\rightarrow \infty$ and
$k\rightarrow \infty$, and then take the limits
$L\rightarrow\infty$ and $\beta_r\rightarrow\infty$. Whenever we
take the limits $L\rightarrow\infty$ and
$\beta_r\rightarrow\infty$ in an LSM of some random matrix
$\mb{A}$ or in the corresponding Stieltjes transform, we denote
the asymptotic expressions by
\begin{align}\nonumber
f_\mb{A}^{(\infty)}(x)\triangleq \lim_{L,\beta_r\rightarrow\infty}
f_\mb{A}(x) \quad\textrm{ and }\quad
G_\mb{A}^{(\infty)}(s)\triangleq \lim_{L,\beta_r\rightarrow\infty}
G_\mb{A}(s).
\end{align}
\begin{theorem2}
Let \mbox{$\mb{H}_1\in\mathbb{C}^{ n_\mr{d}\times k}$},
\mbox{$\mb{H}_2,\ldots,\mb{H}_{L}\in\mathbb{C}^{k\times k}$} and
\mbox{$\mb{H}_{L+1}\in\mathbb{C}^{k\times n_\mr{s}}$} be
independent random matrices with elements of unit-variance
fulfilling the conditions needed for the Mar\v{c}enko-Pastur law
and define $\beta_s\triangleq n_\mr{s}/n_\mr{d}$ and
$\beta_r\triangleq k/n_\mr{d}$. Let \mbox{$\mathsf{snr}$} be a
positive constant. Then, the Stieltjes transform corresponding to
the LSM of
\begin{align}
\mb{C}
&=\mb{R}_s\cdot\mb{R}_n^{-1}=P\alpha^L\cdot\tilde{\mb{R}}_s\cdot\Bigg[\sum_{l=0}^L\alpha^l\mb{R}_{n,l}\Bigg]^{-1},\nonumber\\
\textrm{with }&P=\frac{1-\alpha^{L+1}}{(1-\alpha)\cdot\alpha^L}\cdot\mathsf{snr},\nonumber\\
&\alpha=\frac{P}{1+P},\nonumber\\
&\tilde{\mb{R}}_s\triangleq
\frac{1}{n_\mr{s} k^L}\mb{H}_{1}\cdots
\mb{H}_{L+1}\mb{H}_{L+1}^\mr{H}\cdots
\mb{H}_{1}^\mr{H},\nonumber\\
&\mb{R}_{n,l}\triangleq
\left\{%
\begin{array}{ll}
    \mb{I}_{n_\mr{d}} &, \mathrm{if \hspace{.1cm} } l=0 , \\
     \frac{1}{k^l}\mb{H}_{1}\cdots
\mb{H}_{l}\mb{H}_{l}^\mr{H}\cdots\mb{H}_{1}^\mr{H} &, \mathrm{else
}.\nonumber
\end{array}%
\right.
\end{align}
in the limits \mbox{$\beta_r\rightarrow \infty$},
\mbox{$L\rightarrow \infty$}, but \mbox{$\beta_s$} fixed, converges
to
\begin{align}
G_{\mb{C}}^{(\infty)}(s)&=\left\{%
\begin{array}{ll}
   \frac{1}{\mathsf{snr}}G_\mr{MP}^{\left(\beta_s\right)}\left(\frac{s}{\mathsf{snr}}\right) & \mathrm{\hspace{-.3cm}, if  \hspace{.1cm}} L \in \mathcal{O}\left(\beta_r^{1-\varepsilon}\right), \varepsilon>0,\\
    s^{-1} & \mathrm{\hspace{-.3cm}, if \hspace{.1cm} } L \in
    \Omega(\beta_r^{1+\varepsilon}), \varepsilon>0.
\end{array}%
\right.
\end{align}
Furthermore, if \mbox{$L \in \Theta( \beta_r)$} the Stieltjes
transform converges to some \mbox{$G_{\mb{C}}^{(\infty)}(s)\neq
s^{-1}$} in this limit.
\end{theorem2}
\noindent We have introduced the parameters \mbox{$P$} and
\mbox{$\mathsf{snr}$} such that they correspond to the average
transmit power at the source and the average SNR at the destination.
The case
\mbox{$G_{\mb{C}}^{(\infty)}(s)=\frac{1}{\mathsf{snr}}G_\mr{MP}^{\left(\beta_s\right)}\left(\frac{s}{\mathsf{snr}}\right)$}
corresponds to a point-to-point MIMO channel scenario and thus
generalizes \cite{veniajournal} and \cite{leveque} in the sense that
it gives a condition on how fast the number of relays per layer
needs to grow with the number of hops in order to approach the
Mar\v{c}enko-Pastur law for increasing \mbox{$L$}. The case
\mbox{$G_{\mb{C}}^{(\infty)}(s)=s^{-1}$} can be seen as a
generalization of Theorem 4 in \cite{mueller2}, which states that
almost all eigenvalues vanish, i.e.,
\mbox{$f_\mb{C}^{(\infty)}(x)=\delta(x)$} a.s., if the aspect ratios
\mbox{$\beta_r$} remain finite. The case of linear scaling of
\mbox{$\beta_r$} in \mbox{$L$} constitutes the threshold between the
previous two regimes, but still suffices in order to sustain a
non-zero fraction of the spatial degrees of freedom. In summary, the
theorem thus states that the capacity scales linearly in
$\min\{n_\mr{s},n_\mr{d}\}$ as long as $k$ scales at least linearly
in both $L$ and $\min\{n_\mr{s},n_\mr{d}\}$ and the SNR at the
destination is kept constant.

\section{Asymptotic Analysis}
This section provides the proof of Theorem 2. We start out with
the lemma below, which will allow for inferring a corollary to
Theorem 1. This corollary is the key to the proof of Theorem 2 in
this section. In order not to interrupt the logical flow of the
section two rather technical lemmas required for the proof of
Theorem 2, are stated and proven in the Appendix of this paper.
\begin{lemma1}
For any \mbox{$\varepsilon>0$}, some function
\mbox{$g:\mathbb{R}\longrightarrow \mathbb{R}_+$}, and a positive
constant \mbox{$c$}
\begin{align}\nonumber
\lim_{\kappa\rightarrow\infty}\left(\frac{c}{\kappa}+1\right)^{g(\kappa)}=\left\{
\begin{array}{ll}
    1, &\mr{if} \quad g(\kappa)\in {\cal O}(\kappa^{1-\varepsilon}), \\
    \infty, &\mr{ if } \quad g(\kappa)\in \Omega (\kappa^{1+\varepsilon}).
\end{array}%
\right.
\end{align}
Furthermore, if \mbox{$g(\kappa)\in \Theta( \kappa)$} there exist
constants \mbox{$M_2\geq M_1>0$}, such that
\begin{align}\nonumber
&\liminf_{\kappa\rightarrow\infty}\left(\frac{c}{\kappa}+1\right)^{g(\kappa)}=
    e^{cM_1},\nonumber\\
&\limsup_{\kappa\rightarrow\infty}\left(\frac{c}{\kappa}+1\right)^{g(\kappa)}=
    e^{cM_2}.\nonumber
\end{align}
\end{lemma1}

\noindent {\bf Proof.} The lemma follows from the fact that the
limit can be taken inside a continuous function, which allows as
to write
\begin{align}
\lim_{\kappa\rightarrow\infty}\left(\frac{c}{\kappa}+1\right)^{g(\kappa)}
&=\exp\left(\lim_{\kappa\rightarrow\infty}g(\kappa )\cdot\log
\left(\frac{c}{\kappa}+1\right)\right),\label{eq: 1}
\end{align}
and the rule of Bernoulli-l'Hospital applied to
\mbox{$g(\kappa)=M\kappa^\gamma$}, where $M$ and $\gamma$ are
positive constants:
\begin{align}
\lim_{\kappa\rightarrow\infty} M \kappa^\gamma\cdot\log
\left(\frac{c}{\kappa}+1\right)
&=\frac{cM\kappa^{\gamma}}{\gamma\left(c+\kappa\right)}.\label{eq:
2}
\end{align}
If \mbox{$g(\kappa)\in\mathcal{O}(\kappa^{1-\varepsilon})$}, by
definition there exists some \mbox{$M>0$}, such that the exponent
in \eqref{eq: 1} can be upper-bounded by
\begin{align}\label{eq: blab}
\lim_{\kappa\rightarrow\infty}g(\kappa)\cdot\log
\left(\frac{c}{\kappa}+1\right)&\leq\lim_{\kappa\rightarrow\infty}M\kappa^{1-\varepsilon}\cdot\log
\left(\frac{c}{\kappa}+1\right).
\end{align}
Evaluating \eqref{eq: 2} for $\gamma=1-\varepsilon$ renders this
upper bound zero, which establishes that also the left hand side
(LHS) of \eqref{eq: blab} becomes zero and \eqref{eq: 1} evaluates
to one in this case.

Analogously, if
\mbox{$g(\kappa)\in\Omega(\kappa^{1+\varepsilon})$}, there exists
some $M>0$, such that the exponent in \eqref{eq: 1} can be lower
bounded according to
\begin{align}\label{eq: blabl}
\lim_{\kappa\rightarrow\infty}g(\kappa)\cdot\log
\left(\frac{c}{\kappa}+1\right)&\geq\lim_{\kappa\rightarrow\infty}M\kappa^{1+\varepsilon}\cdot\log
\left(\frac{c}{\kappa}+1\right).
\end{align}
Evaluating \eqref{eq: 2} for $\gamma=1+\varepsilon$ renders the
upper bound infinite, which establishes that both the LHS of
\eqref{eq: blabl} and \eqref{eq: 1} grow without bound in this
case.

Finally, if \mbox{$g(\kappa)\in\Theta(\kappa)$}, there exist
constants $M_1$ and $M_2$, fulfilling \mbox{$M_2\geq M_1>0$}, such
that according to \eqref{eq: 2} evaluated for $\gamma=1$ the
exponent in \eqref{eq: 1} is sandwiched between
\begin{align}
c M_1&=\lim_{\kappa\rightarrow\infty}M_1\kappa\cdot\log
\left(\frac{c}{\kappa}+1\right)\nonumber\\
&\leq\lim_{\kappa\rightarrow\infty}g(\kappa)\cdot\log
\left(\frac{c}{\kappa}+1\right)\nonumber\\
&\leq\lim_{\kappa\rightarrow\infty}M_2 \kappa\cdot\log
\left(\frac{c}{\kappa}+1\right)= c M_2\nonumber,
\end{align}
which establishes the second part of the lemma.
\mbox{$\blacksquare$}

\begin{corollary1}
With the notation and assumptions from Theorem 2 the Stieltjes
transform of the LSM corresponding to \mbox{$\tilde{\mb{R}}_s$} in
the limit \mbox{$k\rightarrow \infty$} and \mbox{$L\rightarrow
\infty$} converges to
\begin{align}\label{eq: signal}
G_{\tilde{\mb{R}}_s}^{(\infty)}(s)=\left\{%
\begin{array}{ll}
    G_\mr{MP}^{\left(\beta_s\right)}(s), &\mr{if} \quad L \in \mathcal{O}
     (\beta_r^{1-\varepsilon})
     , \\
     s^{-1} , &\mr{if} \quad L \in \Omega(\beta_r^{1+\varepsilon}).
\end{array}
\right.
\end{align}
Also, if \mbox{$L\in \Theta (\beta_r)$} the LSM of
\mbox{$\tilde{\mb{R}}_s$} converges to a distribution corresponding
to a Stieltjes transform
\mbox{$G_{\tilde{\mb{R}}_s}^{(\infty)}(s)\neq s^{-1}$} in this
limit.

Furthermore, if  \mbox{$L \in \mathcal{O}(\beta_r^{1-\varepsilon})$}
the Stieltjes transforms of the LSMs corresponding to the
\mbox{$\mb{R}_{n,l}$}'s, for \mbox{$l\in\{1,\ldots,L\}$}, in the
limit \mbox{$k\rightarrow \infty$} and \mbox{$L\rightarrow \infty$}
converge to
\begin{align}\label{eq: noise}
G_{\mb{R}_{n,l}}^{(\infty)}(s)=(s-1)^{-1}.
\end{align}
\end{corollary1}
The part of the corollary related to the $\mb{R}_{n,l}$'s is
actually stated more generally than needed in this paper. In fact,
we are only going to use that for some fixed positive integer
\mbox{$L_\mr{t}<L$}, the LSMs of the $\mb{R}_{n,l}$'s, where
\mbox{$l\in\{1,\ldots,L_\mr{t}\}$}, are given by a Dirac delta at
one, i.e., have the Stieltjes transform \eqref{eq: noise}, when
\mbox{$\beta_r\rightarrow\infty$}, independently of the scaling of
$\beta_r$ in $L$. This is trivially guaranteed by the corollary,
since when \mbox{$L_\mr{t}$} is constant, \mbox{$L_\mr{t} \in
\mathcal{O}(\beta_r^{1-\varepsilon})$} is fulfilled naturally as
$\beta_r\rightarrow\infty$.

\noindent{\bf Proof of Corollary} We treat \eqref{eq: signal} first.
An implicit equation for the Stieltjes transform of the LSM
corresponding to \mbox{$\tilde{\mb{R}}_s$} is given by \eqref{eq:
concatenated stieltjes}, where we set \mbox{$N=L+1$},
\mbox{$\beta_0=1$}, \mbox{$\beta_n=\beta_r$} for
\mbox{$n\in\{1,\ldots N-1\}$}, and \mbox{$\beta_{N}=\beta_s$}
according to our notation:
\begin{align}
\frac{G_{\tilde{\mb{R}}_s}}{\beta_s}\cdot\frac{s
G_{\tilde{\mb{R}}_s}-1+\beta_s}{\beta_r}\cdot&\overbrace{\left(\frac{s
G_{\tilde{\mb{R}}_s}-1+\beta_r}{\beta_r}\right)^{L-1}}^{\Psi_s(\beta_r,L)}\cdot\left(s
G_{\tilde{\mb{R}}_s}-1+\beta_r\right)+s
G_{\tilde{\mb{R}}_s}=1.\label{eq: blabla}
\end{align}
We apply Lemma 1 to \mbox{$\Psi_s(\beta_r,L)$}, where we identify
$\beta_r$ with $\kappa$ and $L$ with $g(\kappa)$. In the limits
\mbox{$\beta_r \rightarrow \infty$}, \mbox{$L\rightarrow \infty$}
and \mbox{$L \in \mathcal{O} (\beta_r^{1-\varepsilon})$} this
yields \mbox{$\Psi_s(\beta_r,L)\rightarrow 1$}. Accordingly,
\eqref{eq: blabla} simplifies to the quadratic equation
\begin{align}\label{eq: blablab}
\beta_s^{-1}s{G_{\tilde{\mb{R}}_s}^{(\infty)}}^2(s)+\left(s+1-\beta_s^{-1}\right)
G_{\tilde{\mb{R}}_s}^{(\infty)}=1.
\end{align}
in this limit. The solution to \eqref{eq: blablab} is the
Stieltjes transform of the Mar\v{c}enko-Pastur law \eqref{eq:
Marcenko} with \mbox{$\beta=\beta_s$}. If \mbox{$L \in
\Omega(\beta_r^{1+\varepsilon})$} we know from Lemma 1 that
\mbox{$\Psi_s(\beta_r,L)$} grows without bounds for
\mbox{$L\rightarrow\infty$}. The numbers of factors in the LHS of
 \eqref{eq: concatenated stieltjes} grows with $L$ in this case. Theorem 4 in reference
\cite{mueller2} states that for \mbox{$N\rightarrow \infty$} the
Stieltjes transform in \eqref{eq: concatenated stieltjes} converges
to \mbox{$G_{\tilde{\mb{R}}_s}^{(\infty)}=s^{-1}$}, if the
\mbox{$\beta_n$} are uniformly bounded. In fact, the conditions
needed for this theorem when
\mbox{$\tilde{\beta}\triangleq\beta_1=\ldots=\beta_{N-1}$}, can be
relaxed to \mbox{$N\in\Omega(\tilde{\beta}^{1+\varepsilon})$}, while
the proof in \cite{mueller2} remains valid. Accordingly, the second
case in \eqref{eq: signal} follows immediately.

For the case \mbox{$L \in \Theta(\beta_r)$} we know from Lemma 1
that
\mbox{$\Psi_s(\beta_r,L)=e^{d(sG_{\tilde{\mb{R}}_s}^{(\infty)}(s)-1)}$}
in the limit of interest for some \mbox{$d>0$}. Thus \eqref{eq:
concatenated stieltjes} simplifies to
\begin{align}\label{eq: 3}
&G_{\tilde{\mb{R}}_s}^{(\infty)}(s)\left(e^{d(sG_{\tilde{\mb{R}}_s}^{(\infty)}(s)-1)}\left(\beta_s^{-1}
s G_{\tilde{\mb{R}}_s}^{(\infty)}+1-\beta_s^{-1}\right)+s\right)=1.
\end{align}
There exists no closed form solution to this implicit equation.
However, it is easily verified that
\mbox{$G_{\tilde{\mb{R}}_s}^{(\infty)}=s^{-1}$} does not satisfy
\eqref{eq: 3}.

For \eqref{eq: noise}, we obtain the equation for the Stieltjes
transform corresponding to the LSM of \mbox{$\mb{R}_{n,l}$}, where
$l\in\{1,\ldots,L\}$, from \eqref{eq: concatenated stieltjes} with
$N=l$, \mbox{$\beta_0=1$} and \mbox{$\beta_n=\beta_r$} for
\mbox{$n\in\{1,\ldots,l\}$} as
\begin{align}\label{eq: 4}
&\frac{G_{\mb{R}_{n,l}}(s)}{\beta_r}\cdot\overbrace{\left(\frac{s
G_{\mb{R}_{n,l}}(s)-1+\beta_r}{\beta_r}\right)^{l-1}}^{\Psi_n(\beta_r,l)}\cdot\left(
sG_{\mb{R}_{n,l}}(s)-1+\beta_r\right) +sG_{\mb{R}_{n,l}}(s)=1.
\end{align}
Let's consider the case \mbox{$l=L$} first. If
\mbox{$L\in\mathcal{O}\left(\beta_r^{1-\varepsilon}\right)$},
$\Psi_n(\beta_r,l)$ converges to one in the limit
$k\rightarrow\infty$ and $L\rightarrow\infty$ by Lemma 1.
Therefore, in this limit \eqref{eq: 4} simplifies to
\begin{align}\label{eq: aa}
G_{\mb{R}_{n,L}}^{(\infty)}(s)+sG_{\mb{R}_{n,L}}^{(\infty)}(s)=1.
\end{align}
The solution to \eqref{eq: aa} is given by
\mbox{$G_{\mb{R}_{n,L}}^{(\infty)}(s)=(s+1)^{-1}$}. The same is
trivially true for all $\mb{R}_{n,l}$ with \mbox{$l<L$}, since
whenever
\mbox{$L\in\mathcal{O}\left(\beta_r^{1-\varepsilon}\right)$}, this
implies that also \mbox{$l\in\mathcal{O}(\beta_r^{1-\varepsilon})$}.
 \mbox{$\blacksquare$}

We are now ready to prove Theorem 2. Besides the above corollary, we
use the Lemmas 2 \& 3, which are stated and proven in the Appendix
of this paper.


\noindent{\bf Proof of Theorem 2.} We go through the different
scaling behaviors of $\beta_r$ with $L$,
 subsequently.

\noindent{\bf Cases \mbox{$L \in
\mathcal{O}\left(\beta_r^{1-\varepsilon}\right)$} and \mbox{$L \in
\Theta\left(\beta_r\right)$}: } Firstly, we show for \emph{both}
these cases that the LSM of $\mb{R}_n$ takes on the shape of a Dirac
delta at some positive constant in the limit of interest. The proof
is based on a truncation of the relay chain between the stages
\mbox{$L_\mr{t}$} and \mbox{$L_{t}-1$}. By choosing
\mbox{$L_\mr{t}$} large enough, we can achieve that the accumulated
noise power originating from the relay stages
\mbox{$L_{t},\ldots,L$} is sufficiently attenuated before it reaches
the destination. More specifically, we claim that for any
\mbox{$\varepsilon>0$} there exist positive integers
\mbox{$L_\mr{t}^{(1)}$} and
\mbox{$n_0^{(1)}(L,L_\mr{t})$}\footnote{We write
\mbox{$n_0^{(1)}(L,L_\mr{t})$} in order to emphasize that $n_0{(1)}$
is a function of $L$ and $L_\mr{t}$}, such that for all \mbox{$
n\geq n_0^{(1)}(L,L_\mr{t})$}, for all
\mbox{$L_\mr{t}>L_\mr{t}^{(1)}$} and \mbox{$L$} arbitrarily large
a.s.\footnote{The Trace norm of the matrix
$\mb{A}\in\mathbb{C}^{n\times n}$ is defined as
\mbox{$\|\mb{A}\|_\mr{Tr}=\sum_{i=1}^n\lambda_i\{\mb{A}\}$}.}
\begin{align}\label{eq: asdfasdf}
&\frac{1}{n_\mr{d}}\left\|\sum_{l=L_\mr{t}+1}^{L}\left(\frac{\alpha}{k}\right)^l
\mb{H}_1\cdots\mb{H}_l\mb{H}_l^\mr{H}\cdots\mb{H}_1^\mr{H}
\right\|_\mr{Tr}<\frac{\varepsilon}{3}.
\end{align}
We prove this by the following chain of inequalities:
\begin{align}
&\frac{1}{n_\mr{d}}\left\|\sum_{l=L_\mr{t}}^{L}\left(\frac{\alpha}{k}\right)^l
\mb{H}_1\cdots\mb{H}_l\mb{H}_l^\mr{H}\cdots\mb{H}_1^\mr{H}
\right\|_\mr{Tr}\nonumber\\
&\hspace{.3cm}\leq \frac{1}{n_\mr{d}}\sum_{l=L_\mr{t}}^{L}\alpha^l
\left\|
\frac{1}{k^l}\mb{H}_1\cdots\mb{H}_l\mb{H}_l^\mr{H}\cdots\mb{H}_1^\mr{H}
\right\|_\mr{Tr}\nonumber\\
&\hspace{.3cm}\leq \max_{l^{'}=L_{t},\ldots,L}\left\{
\frac{1}{n_\mr{d}}\left\|
\frac{1}{k^{l^{'}}}\mb{H}_1\cdots\mb{H}_{l^{'}}\mb{H}_{l^{'}}^\mr{H}\cdots\mb{H}_1^\mr{H}
\right\|_\mr{Tr}\right\}\cdot\sum_{l=L_\mr{t}+1}^{\infty}\alpha^l
\nonumber\\
&\hspace{.3cm}= \max_{l^{'}=L_{t},\ldots,L}\left\{
\frac{1}{n_\mr{d}}\left\|
\frac{1}{k^{l^{'}}}\mb{H}_1\cdots\mb{H}_{l^{'}}\mb{H}_{l^{'}}^\mr{H}\cdots\mb{H}_1^\mr{H}
\right\|_\mr{Tr}\right\}\cdot\frac{\alpha^{L_\mr{t}}}{1-\alpha}.\label{eq:
casec}
\end{align}
In the first step we applied the triangle inequality and used the
homogeneity of the Trace norm. In the second step we upper-bounded
the coefficients of the \mbox{$\alpha^l$}'s by the maximum
coefficient. Afterwards, we let the number of summands go to
infinity, which strictly increases the term, since all added
summands are positive. A standard identity for geometric series
allows for eliminating the sum. \emph{All} arguments of the
\mbox{$\max\{\cdot\}$} function in \eqref{eq: casec} converge to
one a.s.. This follows immediately from Theorem 3 in
\cite{mueller2}. Therefore, we can choose an
\mbox{$n_0^{(1)}(L,L_\mr{t})$} large enough, such that even the
maximum of the \mbox{$L-L_\mr{t}+1$} terms is arbitrarily close to
one a.s. for all \mbox{$n\geq n_0^{(1)}(L,L_\mr{t})$}. In
particular, if \mbox{$L_\mr{t}>\log_\alpha(
(1-\alpha)\cdot\varepsilon/3)$}, we can make
\mbox{$n_0^{(1)}(L,L_\mr{t})$} large enough, such that \eqref{eq:
asdfasdf} is fulfilled a.s. for all \mbox{$n\geq
n_0^{(1)}(L,L_\mr{t})$}.

In a next step we can choose \mbox{$L$} (and thus $\beta_r$) large
enough such that the accumulated noise power originating from the
relays \mbox{$1,\ldots,L_\mr{t}-1$} becomes sufficiently white. This
means, for the fixed \mbox{$L_\mr{t}$} and the same
\mbox{$\varepsilon$} as defined above there exist
\mbox{$L_0>L_\mr{t}$} and $n_0^{(2)}(L,L_\mr{t})$, such that a.s.
for all \mbox{$L\geq L_0$} and for all \mbox{$n\geq
n_0^{(2)}(L,L_\mr{t})$}
\begin{align}\nonumber
&\frac{1}{n_\mr{d}}\left\|\frac{1-\alpha^{L_\mr{t}}}{1-\alpha}\mb{I}_{n_\mr{d}}-\sum_{l=0}^{L_\mr{t}-1}\left(\frac{\alpha}{k}\right)^l
\mb{H}_1\cdots \mb{H}_l\mb{H}_l^\mr{H}\cdots\mb{H}_1^\mr{H}
\right\|_\mr{Tr}<\frac{\varepsilon}{3}.
\end{align}
The proof is similar to the one above:
\begin{align}
&\frac{1}{n_\mr{d}}\left\|\frac{1-\alpha^{L_\mr{t}}}{1-\alpha}\mb{I}_{n_\mr{d}}-\sum_{l=0}^{L_\mr{t}-1}\left(\frac{\alpha}{k}\right)^l
\mb{H}_1\cdots \mb{H}_l\mb{H}_l^\mr{H}\cdots\mb{H}_1^\mr{H}
\right\|_\mr{Tr}\nonumber\\
&\hspace{.3cm}=\frac{1}{n_\mr{d}}\left\|\sum_{l=0}^{L_\mr{t}-1}\alpha^l\cdot\left(\mb{I}_{n_\mr{d}}-\frac{1}{k^l}
\mb{H}_1\cdots \mb{H}_l\mb{H}_l^\mr{H}\cdots\mb{H}_1^\mr{H}\right)
\right\|_\mr{Tr}\nonumber\\
&\hspace{.3cm}\leq
\frac{1}{n_\mr{d}}\sum_{l=0}^{L_\mr{t}-1}\alpha^l
\left\|\mb{I}_{n_\mr{d}}-\frac{1}{k^l} \mb{H}_1\cdots
\mb{H}_l\mb{H}_l^\mr{H}\cdots\mb{H}_1^\mr{H}
\right\|_\mr{Tr}\nonumber\\
&\hspace{.3cm}\leq \max_{l^{'}\in\{1,\ldots,L_\mr{t}-1\}}\left\{
\frac{1}{n_\mr{d}}\left\|\mb{I}_{n_\mr{d}}-\frac{1}{k^l}
\mb{H}_1\cdots \mb{H}_l\mb{H}_l^\mr{H}\cdots\mb{H}_1^\mr{H}
\right\|_\mr{Tr}\right\} \cdot \sum_{l=0}^{\infty}\alpha^l\nonumber\\
&\hspace{.3cm}=\max_{l^{'}\in\{1,\ldots,L_\mr{t}-1\}}\left\{
\frac{1}{n_\mr{d}}\left\|\mb{I}_{n_\mr{d}}-\frac{1}{k^l}
\mb{H}_1\cdots \mb{H}_l\mb{H}_l^\mr{H}\cdots\mb{H}_1^\mr{H}
\right\|_\mr{Tr}\right\}\cdot\frac{1}{1-\alpha}.\label{eq: cased}
\end{align}
Again, the first identity is a standard identity for a geometric
series. In this case the convergence of the arguments of the
\mbox{$\max(\cdot)$} function is guaranteed by Corollary 1 and Lemma
2 (refer to Appendix): \mbox{Corollary 1} tells us that the LSMs of
all the \mbox{$\frac{1}{k^l} \mb{H}_1\cdots
\mb{H}_l\mb{H}_l^\mr{H}\cdots\mb{H}_1^\mr{H}$}'s, where
\mbox{$l\in\{1,\ldots,L_\mr{t}-1\}$}, converge to a Dirac at one in
the limit under consideration. Knowing this, \mbox{Lemma 2}
guarantees us that the respective Trace norms go to zero. Thus,
there exist $L_0$ and \mbox{$n_0^{(2)}(L,L_\mr{t})$}, such that a.s.
the maximum of the \mbox{$L_\mr{t}-1$} terms is small enough to make
\eqref{eq: cased} smaller than \mbox{$\varepsilon/3$} for all
\mbox{$L>L_0$} and \mbox{$n\geq n_0^{(2)}(L,L_\mr{t})$}.

With the choices \mbox{$L_\mr{t}=\max\{L_\mr{t}^{(1)},\log_\alpha(
(1-\alpha)\cdot\varepsilon/3)\}$} and
\mbox{$n_0(L,L_\mr{t})=\max\{n_0^{(1)}(L,L_\mr{t}),n_0^{(2)}(L,L_\mr{t})\}$},
we can finally conclude by the triangle inequality that for all
\mbox{$L>L_0$} and \mbox{$n\geq n_0(L,L_\mr{t})$} a.s.
\begin{align}
&\frac{1}{n_\mr{d}}\left\|\frac{1-\alpha^{L+1}}{1-\alpha}\cdot\mb{I}_{n_\mr{d}}-\mb{R}_n
\right\|_\mr{Tr}=\frac{1}{n_\mr{d}}\left\|\frac{1-\alpha^{L+1}}{1-\alpha}\cdot\mb{I}_{n_\mr{d}}-
\sum_{l=0}^L\alpha^l\mb{R}_{n,l}
\right\|_\mr{Tr}\nonumber\\
&\hspace{.3cm}=\frac{1}{n_\mr{d}}\left\|\frac{1-\alpha^{L+1}}{1-\alpha}\cdot\mb{I}_{n_\mr{d}}-\sum_{l=0}^{L}\left(\frac{\alpha}{k}\right)^l
\mb{H}_1\cdots \mb{H}_l\mb{H}_l^\mr{H}\cdots\mb{H}_1^\mr{H}
\right\|_\mr{Tr}\nonumber\\
&\hspace{.3cm}=\frac{1}{n_\mr{d}}\left\|\frac{\alpha^{L_\mr{t}}-\alpha^{L}}{1-\alpha}\cdot\mb{I}_{n_\mr{d}}+\frac{1-\alpha^{L_\mr{t}}}{1-\alpha}\cdot\mb{I}_{n_\mr{d}}-\sum_{l=0}^{L_\mr{t}-1}\left(\frac{\alpha}{k}\right)^l
\mb{H}_1\cdots
\mb{H}_l\mb{H}_l^\mr{H}\cdots\mb{H}_1^\mr{H}-\sum_{l=L_\mr{t}}^{L}\left(\frac{\alpha}{k}\right)^l
\mb{H}_1\cdots \mb{H}_l\mb{H}_l^\mr{H}\cdots\mb{H}_1^\mr{H}
\right\|_\mr{Tr}\nonumber\\
&\hspace{.3cm}\leq\frac{1}{n_\mr{d}}\left\|\frac{\alpha^{L_\mr{t}}}{1-\alpha}\cdot\mb{I}_{n_\mr{d}}
\right\|_\mr{Tr}+\frac{1}{n_\mr{d}}\left\|\frac{1-\alpha^{L_\mr{t}}}{1-\alpha}\cdot\mb{I}_{n_\mr{d}}-\sum_{l=0}^{L_\mr{t}-1}\left(\frac{\alpha}{k}\right)^l
\mb{H}_1\cdots \mb{H}_l\mb{H}_l^\mr{H}\cdots\mb{H}_1^\mr{H}
\right\|_\mr{Tr}\nonumber\\
&\hspace{.3cm}\hspace{.3cm}+\frac{1}{n_\mr{d}}\left\|\sum_{l=L_\mr{t}}^{L}\left(\frac{\alpha}{k}\right)^l
\mb{H}_1\cdots \mb{H}_l\mb{H}_l^\mr{H}\cdots\mb{H}_1^\mr{H}
\right\|_\mr{Tr}<\frac{\varepsilon}{3}+\frac{\varepsilon}{3}+\frac{\varepsilon}{3}
=\varepsilon.\nonumber
\end{align}
Here, we decomposed the terms in a way that allowed us to obtain a
sum of precisely the expressions we had proven to converge to zero
before. Besides standard steps we used the fact that that
\mbox{$|\alpha^{L_\mr{t}}-\alpha^L|<|\alpha^{L_\mr{t}}|$}, since
$L>L_t$ and $\alpha<1$. By Lemma 2 we have established that the LSM
of \mbox{$\mb{R}_n$} converges to
\begin{align}\label{eq: xxx}
f_{\mb{R}_n}^{(\infty)}(x)=\delta
\left(x-\frac{1-\alpha^{L+1}}{1-\alpha}\right).
\end{align}
Note that the fact that \emph{almost all} eigenvalues of the
single \mbox{$\mb{R}_{n,l}$}'s are \emph{arbitrarily close} to one
each, does not immediately imply that this is also the case for a
weighted sum of these matrices. This is due to the fact that for
the matrix series \mbox{$\mb{R}_{n,l}$}, where \mbox{$l\in
\{1,\ldots L_\mr{t}\}$}, the identity
\mbox{$\lambda_k\{\sum_{l=0}^{L}\mb{R}_{n,l}\}=
\sum_{l=0}^{L}\lambda_k\{\mb{R}_{n,k}\}$} is fulfilled, if and
only if \emph{all} these eigenvalues are \emph{exactly} equal to
one. This easy attempt of proving \eqref{eq: xxx} must therefore
fail. Also note that the \mbox{$\mb{R}_{n,l}$}'s are \emph{not}
asymptotically free, which prohibits arguing based on the
respective R-transforms \cite{muller-random}.

Since the eigenvalues of the corresponding inverse are the inverse
eigenvalues, i.e.,
\mbox{$\lambda_k\{\mb{R}_n^{-1}\}=\lambda_k^{-1}\{\mb{R}_n\}$}, we
conclude that the LSM of \mbox{$\mb{R}_n^{-1}$} is given by
\mbox{$f_{\mb{R}_n^{-1}}^{(\infty)}(x)=\delta(x-(1+\alpha)/(1-\alpha^{L+1}))$}.
Thus, by Lemma 3 (refer to Appendix) and the respective variable
transformation applied to
\mbox{$f_{\mb{R}_n^{-1}}^{(\infty)}(\cdot)$} the EED of
\mbox{$\mb{C}=\mathsf{snr}\cdot\tilde{\mb{R}}_s\mb{R}_n^{-1}$}
coincides with the EED of
\mbox{$\mathsf{snr}\cdot\tilde{\mb{R}}_s$}, i.e.,
\begin{align}\nonumber
f_{\mb{C}}^{(\infty)}(x)=\frac{1}{\textsf{snr}}f_{\tilde{\mb{R}}_s}\left(\frac{x}{\textsf{snr}}\right)\quad\textrm{
and }\quad
G_{\mb{C}}^{(\infty)}(s)=\frac{1}{\textsf{snr}}G_{\tilde{\mb{R}}_s}\left(\frac{s}{\textsf{snr}}\right).
\end{align}

 By
\mbox{Corollary 1} the LSM of \mbox{$\tilde{\mb{R}}_s$} is given
by the Mar\v{c}enko-Pastur law, in the case that
\mbox{$L\in\mathcal{O}(\beta_r^{1-\varepsilon})$}, which
establishes the first case. In the case \mbox{$L \in
\Theta\left(\beta_r\right)$} a non-zero fraction of the
eigenvalues of \mbox{$\tilde{\mb{R}}_s$} remains non-zero as
\mbox{$L\rightarrow \infty$} by \mbox{Corollary 1}, i.e.,
\mbox{$G_{\tilde{\mb{R}_s}}(s)=G_\mb{C}(s)\neq s^1$}. 

\noindent{\bf Case \mbox{$L \in
\Omega\left(\beta_r^{1+\varepsilon}\right)$}:} This case follows
immediately by Corollary 1. Since asymptotically almost all
eigenvalues of \mbox{$\tilde{\mb{R}}_s$} vanish, also almost all
eigenvalues of \mbox{$\mb{C}=\mb{R}_s\mb{R}_n^{-1}$} need to
approach zero. We rely on the reader's intuition here, that noise
cannot recover degrees of freedom. A formal proof goes along the
lines of the proof of Lemma 3, where $\mb{A}$ is identified with
$\mb{R}_n^{-1}$ and $\mb{B}$ with $\mb{R}_s$. In the end one can
show that the Shannon transforms of the LSMs
$f_{\mb{R}_s\mb{R}_n^{-1}}(\cdot)$ and $f_{\mb{R}_s}(\cdot)$
coincide at a Dirac delta at zero in the limit $L\rightarrow\infty$
and $\beta_r\rightarrow\infty$. \mbox{$\blacksquare$}

\section{Simulation Results}
The above results provide no evidence about speed of convergence.
Since speed of convergence results are generally hard to obtain in
a large matrix dimensions analysis, we resort to a numerical
demonstration for this purpose. In doing so, we specify the
distribution of the elements of $\mb{H}_1,\ldots,\mb{H}_{L+1}$ as
circularly symmetric complex Gaussian with zero-mean and
unit-variance. Recall that all assumptions imposed on this
distribution in Theorem 2 were just related to its first and
second moments. We fix the number of source and destination
antennas to $n\triangleq n_\mr{s}=n_\mr{d}=10$ and plot the
normalized ergodic capacity $C_0=(L+1)\cdot C/n$ as obtained
through Monte Carlo simulations versus the number of relay
clusters, $L$, for an average SNR of 10 dB at the destination. The
number of relays per cluster evolves with the number of clusters
according to \mbox{$k=L^{\gamma}$}, where we vary $\gamma$ between
0 and 3. In principle, one could use the recursive formula for the
Stieltjes transform of $\mb{C}$ obtained in \cite{leveque} rather
than Monte Carlo simulations for generating these plots. However,
the respective evaluations are handy for small $L$ only.

\begin{figure}
        \centering

        \includegraphics[width=\columnwidth]{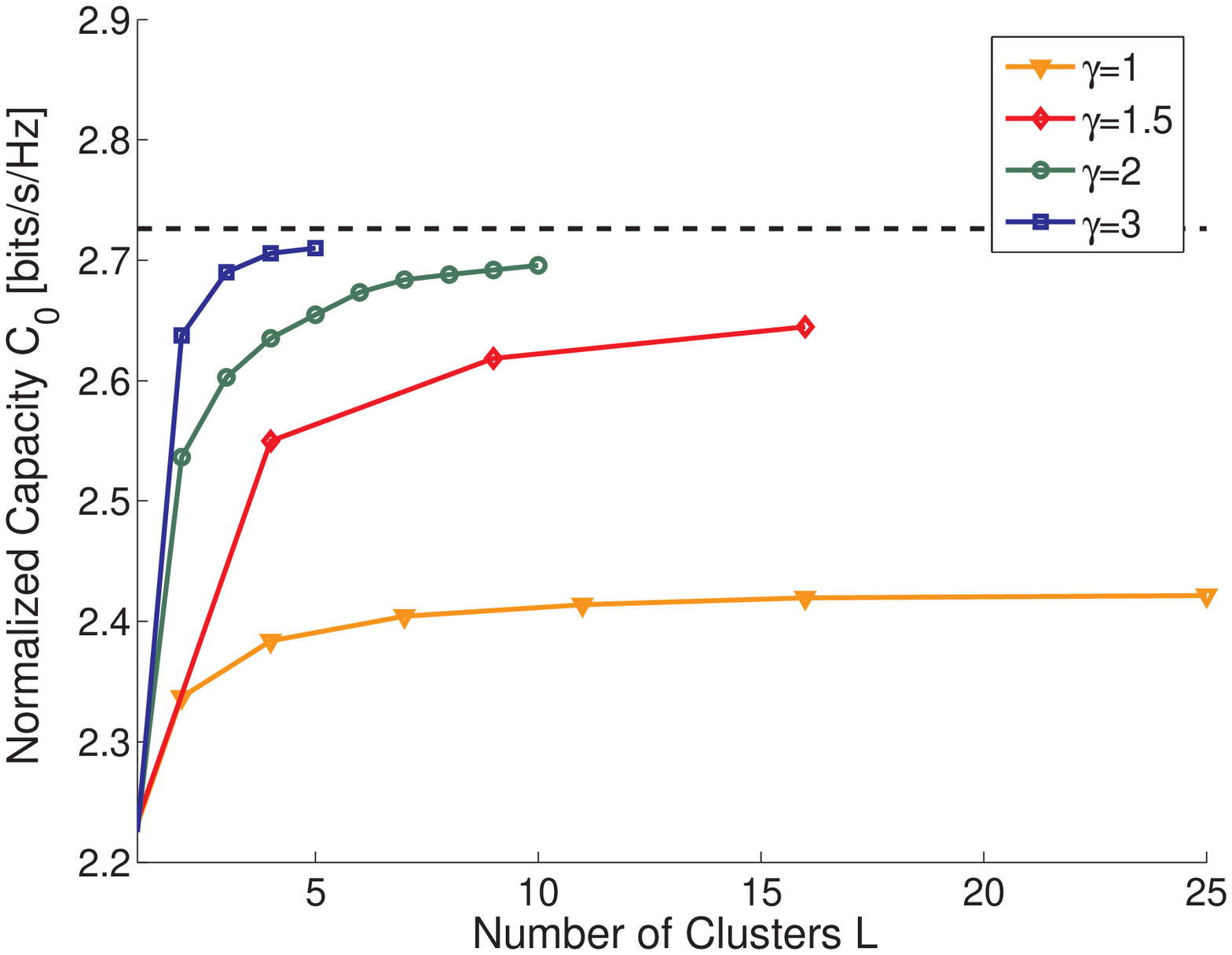}
        \caption{Normalized capacity $C_0$ versus the number of relay clusters $L$ as obtained through Monte Carlo simulations. The number of source and destination antennas is $n=10$. The SNR at the destination is 10 dB. The number of relays per cluster $k$ evolves according to $k=n\cdot L^{\gamma}$. The dashed curve shows the normalized point-to-point MIMO capacity as a reference. }\label{fig: large}
\end{figure}

\begin{figure}
        \centering

        \includegraphics[width=\columnwidth]{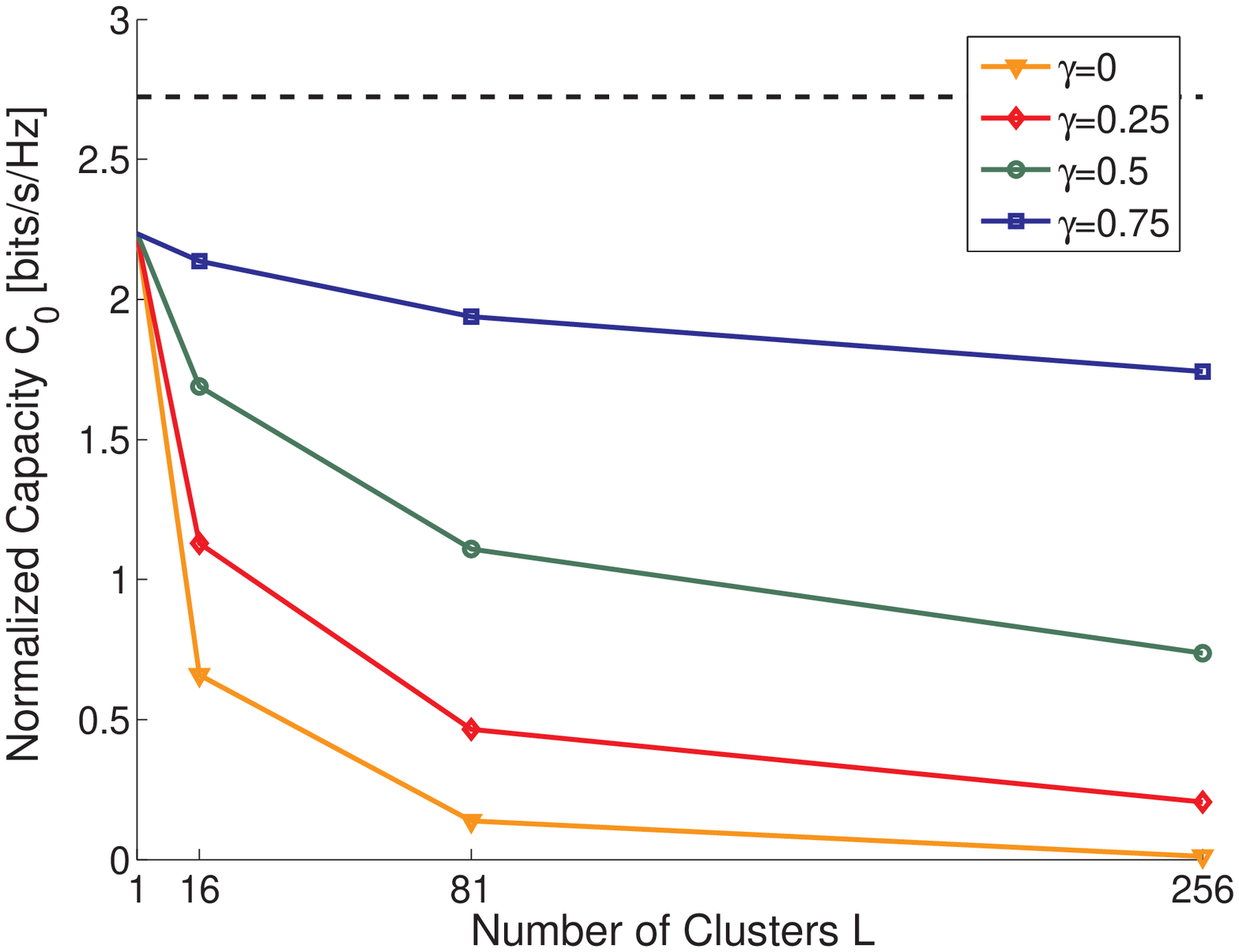}
        \caption{Normalized capacity $C_0$ versus the number of relay clusters $L$ as obtained through Monte Carlo simulations. The number of source and destination antennas is $n=10$. The SNR at the destination is 10 dB. The number of relays per cluster $k$ evolves according to $k=n\cdot L^{\gamma}$. The dashed curve shows the normalized point-to-point MIMO capacity as a reference. }
        \label{fig: small}
\end{figure}

Fig. \ref{fig: large} shows the case of linear and faster scaling
of $k$ in $n$. For $\gamma=1$ the curve flattens out quickly, and
converges to some constant which is smaller than the normalized
point-to-point MIMO capacity, but non-zero, as expected.
Furthermore, we observe that the point-to-point limit is
approached the sooner the bigger $\gamma$ is chosen. For
$\gamma=3$ this is the case after less than 10 hops already. Fig.
\ref{fig: small} shows the case of less than linear scaling of $k$
in $n$. While $C_0$ decreases rather rapidly for constant relay
numbers ($\gamma=0$), we observe that already a moderate growth of
$k$ with $L$ slows down the capacity decay significantly. For
$\gamma=0.5$ an almost threefold capacity gain over the $\gamma=0$
case is achieved for $L=16$. For $\gamma=0.75$ the decay is
tolerable even for very large $L$. Note that for $L=81$ and
$\gamma=0.75$ only 27 relays per layer are used in contrast to the
81 relays needed for linear scaling.

\section{Conclusion}
We have given a criterion how the number of relays per stage needs
to be increased with the number of hops in order to sustain a
non-zero fraction of the spatial degrees of freedom in a MIMO
amplify-and-forward multi-hop network, i.e., linear capacity scaling
in $\min\{n_\mr{s},n_\mr{d}\}$. The necessary and sufficient
condition is an \emph{at least linear} scaling of the relays per
stage in the number of hops.

\section{Acknowledgement}
We gratefully acknowledge the very helpful comments by Veniamin I.
Morgenshtern.

\begin{appendix}
\begin{lemma2}
A random matrix \mbox{$\mb{A}\in\mathbb{C}^{n\times n}$} fulfilling
\mbox{$\lim_{n\rightarrow\infty}n^{-1}\mr{Tr}(\mb{A})=1$} converges
to the identity matrix a.s. in the sense that
\begin{align}\nonumber
\lim_{n\rightarrow\infty}
\frac{1}{n}\|\mb{I}_n-\mb{A}\|_\mr{Tr}=0,
\end{align}
if and only if its EED \mbox{$F_\mb{A}^{n}(x)$} converges to
\mbox{$\sigma(x-1)$}, i.e., a.s.
\begin{align}\nonumber
\lim_{n\rightarrow
\infty}\sup_{x\in\mathbb{R}}|F_\mb{A}^{n}(x)-\sigma(x-1)| =0.
\end{align}
\end{lemma2}

\noindent {\bf Proof.} The lemma follows by a the following
identities:
\begin{align}\nonumber
&\lim_{n\rightarrow \infty}\frac{1}{n}\|\mb{I}_n-\mb{A}\|_\mr{Tr}=
\lim_{n\rightarrow \infty}\frac{1}{n}\sum_{i=1}^n
|1-\lambda_i\{\mb{A}\}|\nonumber\\
&=\lim_{n\rightarrow
\infty}\frac{1}{n}\sum_{i:\lambda_i\{\mb{A}\}\leq 1} \left(1-\lambda_i\{\mb{A}\}\right)  +\frac{1}{n}\sum_{i:\lambda_i\{\mb{A}\}>1}\left(\lambda_i\{\mb{A}\}-1\right)\label{eq: arrange}\\
&=\lim_{n\rightarrow
\infty}\int_0^1 |F_\mb{A}^{(n)}(x)|\cdot dx +\int_1^\infty |F_\mb{A}^{(n)}(x)-1|\cdot dx \nonumber\\
&=\lim_{n\rightarrow \infty}\int_0^\infty
|F_\mb{A}^{(n)}(x)-\sigma(x-1)|\cdot dx=0\label{eq: norm1}\\
\Longleftrightarrow &\lim_{n\rightarrow \infty}\sup_{x\in\mathbb{R}}
|F_\mb{A}^{(n)}(x)-\sigma(x-1)|=0\label{eq: norm2}.
\end{align}
In \eqref{eq: arrange} we arrange the individual summands such that
they can be related to the EED of $\mb{A}$. The equivalence of the
norms in \eqref{eq: norm1} and \eqref{eq: norm2} is established as
follows: For the forward direction consider
\mbox{$\epsilon(x)\triangleq |F^{(n)}(x)-\sigma(x-1)|$} for
\mbox{$x\in[0,1)$}, i.e., \mbox{$\epsilon(x)=|F^{(n)}(x)|$}. Choose
any \mbox{$\Delta\in[-1,0)$}. Since \mbox{$\epsilon(x)$} is
monotonically increasing on the interval of interest, we can write
\begin{align}\nonumber
\int_{1-\Delta}^1|F_\mb{A}^{(n)}(x)-\sigma(x-1)|\cdot
dx>|\Delta|\cdot\epsilon(1+\Delta).
\end{align}
Thus, if \mbox{$\epsilon(1+\Delta)$} does not go to zero for all
\mbox{$\Delta$}, the integral norm cannot go to zero. The same
reasoning can be applied for the interval
\mbox{$\Delta\in[1,\infty)$}.

For the backward part we break the integration in \eqref{eq: norm1}
into two parts. The first integral is from zero to some constant
\mbox{$d>1$}. This part is a Riemann integral over a function that
converges uniformly by \eqref{eq: norm2}. It goes to zero by taking
the limit inside the integral. The second part of the integral is
from \mbox{$d$} to \mbox{$\infty$}. Here, the limit cannot be taken
inside the integral in general. However, we can write
\begin{align}\label{eq: asdfasdfasdf}
\lim_{n\rightarrow\infty}\int_d^\infty 1-F_\mb{A}^{(n)}(x)\cdot
dx=\lim_{n\rightarrow\infty}\int_0^\infty 1-F_\mb{A}^{(n)}(x)\cdot
dx-\lim_{n\rightarrow\infty}\int_0^d 1-F_\mb{A}^{(n)}(x)\cdot
dx=1-1=0.
\end{align}
The first integral on the the right hand side (RHS) converges to one
by the following chain of identities:
\begin{align}
\lim_{n\rightarrow\infty}\int_0^\infty 1-F_\mb{A}^{(n)}(x)\cdot dx
&=\lim_{n\rightarrow\infty}\int_0^\infty 1-\frac{1}{n}\sum_{i=1}^n
1\{\lambda_i<x\}\cdot
dx\nonumber\\
&=\lim_{n\rightarrow\infty}\frac{1}{n}\sum_{i=1}^{n}\int_0^\infty
1-1\{\lambda_i<x\}\cdot
dx\nonumber\\
&=\lim_{n\rightarrow\infty}\frac{1}{n}\sum_{i=1}^n\lambda_i\{\mb{A}\}=\lim_{n\rightarrow\infty}\frac{1}{n}\mr{Tr}[\mb{A}]=1.\nonumber
\end{align}
The second term on the RHS of \eqref{eq: asdfasdfasdf} is identified
to converge to one by taking the limit inside the integral. Again,
this can be done, since we deal with a Riemann integral over a
uniformly convergent function. \mbox{$\blacksquare$}


%
%
%

\begin{lemma3}
Let \mbox{$\mb{A}, \mb{B}\in\mathbb{C}^{n\times n}$} be
positive-semidefinite random matrices with LSMs
\mbox{$f_\mb{A}(x)=\delta(x-1)$} and \mbox{$f_\mb{B}(x)=\psi(x)$},
respectively. Then, the LSM of \mbox{$\mb{A}\mb{B}$} is given by
\mbox{$f_{\mb{A}\mb{B}}(x)=\psi(x)$}.
\end{lemma3}

\noindent {\bf Proof.} We separate the eigenvalues
\mbox{$\mu_i\triangleq\lambda_i\{\mb{A}-\mb{I}_n\}$},
\mbox{$i\in\{1,\ldots,n\}$}, into two sets \mbox{$\mathcal{L}_1$}
and \mbox{$\mathcal{L}_2$}. For a fixed \mbox{$\varepsilon > 0$} the
eigenvalues in the first set fulfill \mbox{$|\mu_i|\leq
\varepsilon$}. The second set contains the eigenvalues which fulfill
\mbox{$|\mu_i|>\varepsilon$}. Firstly, we show that the eigenvalues
in \mbox{$\mathcal{L}_2$} do not have any impact on the LSM of
\mbox{$\mb{A}\mb{B}$}. Since \mbox{$\mb{A}-\mb{I}_n$} is Hermitian,
we can write with \mbox{$\tilde{\mb{A}}\triangleq
\mb{I}_n+\sum_{i:\mu_i\in
\mathcal{L}_1}\mu_i\mb{v}_i\mb{v}_i^\mr{H}$}
\begin{align}\nonumber
\mb{A}\mb{B}&=\tilde{\mb{A}}\mb{B}+\sum_{i:\mu_i\in
\mathcal{L}_2}\mu_i\mb{v}_i\mb{v}_i^\mr{H}\mb{B},
\end{align}
where \mbox{$\mb{v}_i$} denotes the eigenvector corresponding to
\mbox{$\mu_i$}.
The EED of \mbox{$\mb{A}$} a.s. converges to \mbox{$\sigma(x-1)$}.
Therefore, the number of eigenvalues in \mbox{$\mathcal{L}_2$}
grows less than linearly in \mbox{$n$}.
Since the $\mb{v}_i\mb{v}_i^\mr{H}$'s are unit rank matrices, we
conclude that the fraction of differing eigenvalues of
\mbox{$\mb{A}\mb{B}$} and \mbox{$\tilde{\mb{A}}\mb{B}$} goes to zero
as \mbox{$n\rightarrow\infty$}. Thus, they also share the same LSM.

Secondly, we show that the LSM of
\mbox{${\tilde{\mb{A}}}^{-1}+\rho\mb{B}$} is given by
\begin{align}\label{eq: xxy}
f_{{\tilde{\mb{A}}}^{-1}+\rho\mb{B}}(x)=\rho^{-1}f_\mb{B}(\rho^{-1}x-1).
\end{align}
Note that the eigenvalues of \mbox{${\tilde{\mb{A}}}^{-1}$} are the
inverse eigenvalues of \mbox{${\tilde{\mb{A}}}$}. Therefore, we can
write \mbox{${\tilde{\mb{A}}}^{-1}=\mb{I}_n+\mb{\Delta}$}, where
a.s. for any \mbox{$\delta>0$} there exists an $n_0$ such that for
all $n\geq n_0$
\begin{align}\label{eq: xx}
\max_{i\in\{1,\ldots,n\}}\lambda_i\{\mb{\Delta}\}<\delta.
\end{align}
Let's denote the normalized eigenvectors of
\mbox{${\tilde{\mb{A}}}^{-1}+\rho\mb{B}$} corresponding to
\mbox{$\lambda_i\{{\tilde{\mb{A}}}^{-1}+\rho\mb{B}\}$} by
\mbox{$\mb{u}_i$}. By the definition of an eigenvector, we can write
\begin{align}\nonumber
\left(\mb{I}_n+\mb{\Delta}+\rho\mb{B}-\lambda_i\{{\tilde{\mb{A}}}^{-1}+\rho\mb{B}\}\cdot\mb{I}_n\right)\cdot\mb{u}_i=\mb{0}
\end{align}
for \mbox{$i\in\{1,\ldots,n\}$}. Taking \mbox{$\mb{\Delta}\mb{u}_i$}
to the RHS and taking the Eukledian norm yields
\begin{align}\label{eq:bla}
\|\rho\mb{B}\mb{u}_i+(1-\lambda_i\{{\tilde{\mb{A}}}^{-1}+\rho\mb{B}\})\mb{u}_i\|=\|\mb{\Delta}\mb{u}_i\|.
\end{align}
By \eqref{eq: xx} and \mbox{$\|\mb{u}_i\|=1$}, we conclude that for
for all $n\geq n_0$ a.s. also \begin{align}
\|\mb{\Delta}\mb{u}_i\|<\max_{i\in\{1,\ldots,n\}}\lambda_i\{\mb{\Delta}\}<\delta.
\end{align}
Thus, for all \mbox{$i$} the RHS of \eqref{eq:bla} goes to zero as
\mbox{$n\rightarrow\infty$}. Accordingly, we conclude for the LHS
that \mbox{$\mb{u}_i\rightarrow \mb{w}_i$} and
\mbox{$\lambda_i\{{\tilde{\mb{A}}}^{-1}+\rho\mb{B}\}\rightarrow
1+\rho\nu_i$}, where \mbox{$\nu_i$} and \mbox{$\mb{w}_i$} are the
\mbox{$i$}th eigenvalue and eigenvector of \mbox{$\mb{B}$}.
The respective variable transformation in the LSM of
\mbox{$\mb{B}$} yields \eqref{eq: xxy}.

We complete the proof by showing that the Shannon transforms
\cite{shannontransform,rmtbook} of
\mbox{$f_{\tilde{\mb{A}}\mb{B}}(\cdot)$} and
\mbox{$f_{\mb{B}}(\cdot)$} coincide. Note that the Shannon transform
contains the full information about the corresponding distribution.
Consider the quantity \mbox{$\xi\triangleq
 n^{-1}\log\det\left(\tilde{\mb{A}}^{-1}+\rho\mb{B}\right)-n^{-1}\log\det\left(\tilde{\mb{A}}^{-1}\right)$}.
As \mbox{$n\rightarrow \infty$} this quantity converges to the
Shannon transform of \mbox{$f_{\mb{B}}(\cdot)$},
$\Upsilon_\mb{B}(\rho)$,  a.s.:
\begin{align}\nonumber
\lim_{n\rightarrow\infty}\xi&=\int_0^\infty \log x\cdot
f_{\tilde{\mb{A}}^{-1}+\rho\mb{B}}(x)\cdot dx-\int_0^\infty
\log x \cdot f_{\tilde{\mb{A}}^{-1}}(x)\cdot dx\nonumber\\
&= \int_0^\infty \log(1+\rho x)\cdot f_{\mb{B}}(x)\cdot
dx-\int_0^\infty \log x \cdot \delta(x-1)\cdot dx
\nonumber\\
&= \int_0^\infty \log(1+\rho x)f_{\mb{B}}(x)\cdot dx \triangleq
\Upsilon_\mb{B}(\rho)\nonumber
\end{align}
Rewriting
\mbox{$\xi=n^{-1}\log\det\left(\mb{I}_n+\rho\tilde{\mb{A}}\mb{B}\right)$},
we see that \mbox{$\xi$} also converges to the Shannon transform of
\mbox{$f_{\tilde{\mb{A}}\mb{B}}(\cdot)$},
$\Upsilon_{\tilde{\mb{A}}\mb{B}}(\rho)$, a.s.:
\begin{align}\nonumber
\lim_{n\rightarrow\infty}\xi&=\int_0^\infty \log x
f_{\mb{I}_n+\rho\tilde{\mb{A}}\mb{B}}(x)\cdot dx= \int_0^\infty
\log(1+\rho x)f_{\tilde{\mb{A}}\mb{B}}(x)\cdot dx\triangleq
\Upsilon_{\tilde{\mb{A}}\mb{B}}(\rho).\nonumber
\end{align}
Accordingly, we conclude that
\mbox{$f_{\mb{A}\mb{B}}(x)=f_{\tilde{\mb{A}}\mb{B}}(x)=
f_{\mb{B}}(x)=\psi(x)$}. \mbox{$\blacksquare$}
\end{appendix}

\newpage
\bibliography{refs}
\bibliographystyle{ieeetran}

\newpage

\newpage

\newpage

\end{document}